\documentclass{article}
\usepackage{spconf,amsmath,graphicx}
\usepackage{amsmath}
\usepackage{amssymb}
\usepackage{amsfonts}
\usepackage{booktabs}
\usepackage{colortbl}
\usepackage{multirow}
\usepackage{hyperref}
 
\usepackage{todonotes}
 
\usepackage{color}

\usepackage{enumitem}
\setlist{nosep, leftmargin=14pt}
 
\usepackage{mwe}

\title{Uncertainty Estimation in Contrast-Enhanced MR Image Translation with Multi-Axis Fusion}
 
 \name{Ivo~M.~Baltruschat, Parvaneh~Janbakhshi, Melanie~Dohmen, Matthias~Lenga}
 \address{Bayer AG, M\"ullerstr. 178, 13353 Berlin, Germany}
\begin{document}
\maketitle
\begin{abstract}
In recent years, deep learning has been applied to a wide range of medical imaging and image processing tasks. In this work, we focus on the estimation of epistemic uncertainty for 3D medical image-to-image translation. We propose a novel model uncertainty quantification method, Multi-Axis Fusion (MAF), which relies on the integration of complementary information derived from multiple views on volumetric image data. The proposed approach is applied to the task of synthesizing contrast enhanced T1-weighted images based on native T1, T2 and T2-FLAIR scans. The quantitative findings indicate a strong correlation ($\rho_{\text healthy} = 0.89$) between the mean absolute image synthetization error and the mean uncertainty score for our MAF method. Hence, we consider MAF as a promising approach to solve the highly relevant task of detecting synthetization failures at inference time.
\end{abstract}
 
\section{Introduction}
\label{sec:intro}

Deep learning is evolving at light speed and is being applied to a wide range of medical imaging and image processing tasks. However, deep learning models are often considered black boxes, and no measure of uncertainty accompanies their predictions. This is especially problematic in medical image-to-image translation, where the consequences of a wrong target domain image synthetization can be as severe as tumor removal or hallucination \cite{cohenDistributionMatchingLosses2018}. Therefore, the development of methods to estimate the uncertainty of deep learning models is crucial.
Uncertainty in deep learning can be categorized into two types: epistemic and aleatoric uncertainty. Aleatoric uncertainty characterizes inherent noise in observations (i.e., data). Epistemic uncertainty refers to a lack of knowledge of the model (i.e., domain shift, underfitting, etc.) \cite{zouReviewUncertaintyEstimation2023}. In this work, we focus on the estimation of epistemic uncertainty.
 
Bayesian neural networks facilitate the quantification of epistemic uncertainty, but are computationally intensive and complex to train, limiting their practical use \cite{magrisBayesianLearningNeural2023,nealBayesianLearningNeural1996}.
Due to this, multiple methods have been proposed to estimate epistemic uncertainty in deep learning models. Two of the most popular methods are Monte Carlo Dropout (MC-Dropout) \cite{galDropoutBayesianApproximation2016} and Deep Ensemble \cite{lakshminarayananSimpleScalablePredictive2017}. MC-Dropout is a computationally feasible approximation to Bayesian neural networks, providing uncertainty estimates for models' confidence scores. Deep Ensemble is a non-Bayesian solution that yields well-calibrated predictive uncertainty estimates with minimal hyperparameter tuning at the cost of requiring multiple models training. MC-Dropout and Deep Ensemble techniques have shown promising results in uncertainty estimation for image classification and regression tasks\cite{zouReviewUncertaintyEstimation2023}. However, the application of MC-Dropout and Deep Ensemble in image-to-image translation for contrast-enhanced MRI is still relatively unexplored.
In \cite{kendallWhatUncertaintiesWe2017}, a Bayesian deep learning framework that combines both uncertainty types was proposed and had inspired several works in medical image-to-image translation~\cite{zhangReducingUncertaintyUndersampled2019,kleesiekCanVirtualContrast2019, upadhyayUncertaintyawareGANAdaptive2021}. They model the predictive distributions using the Gaussian or a generalized Gaussian distribution. Often, those methods require extensive hyperparameter tuning to balance the estimation of the distribution parameters, e.g., the predictive variance versus the predictive mean.
 
The closest to our method is test-time data augmentation for uncertainty estimation. In \cite{ayhanTesttimeDataAugmentation2022}, Ayhan {\it et al.} proposed to generate augmented examples per test case using simple and randomized transformations and feeding them to a neural network. In this way, samples of an approximate predictive distribution are obtained which can then be used to evaluate the predictive uncertainty.
 
We propose Multi-Axis Fusion \cite{baltruschatScalingUnetSegmentation2021} to estimate the uncertainty for 3D medical image-to-image translation. This method uses the unique properties of $2$D or $2.5$D slice-wise processing of $3$D medical images. Multi-axis fusion was proposed for segmentation and combines predictions of multiple axes~\cite{baltruschatScalingUnetSegmentation2021}. Here, we also use this technique for $3$D medical image-to-image translation and have adapted it for uncertainty estimation.
 
\begin{figure*}[t]
  \centering
  \includegraphics[width=0.95\linewidth]{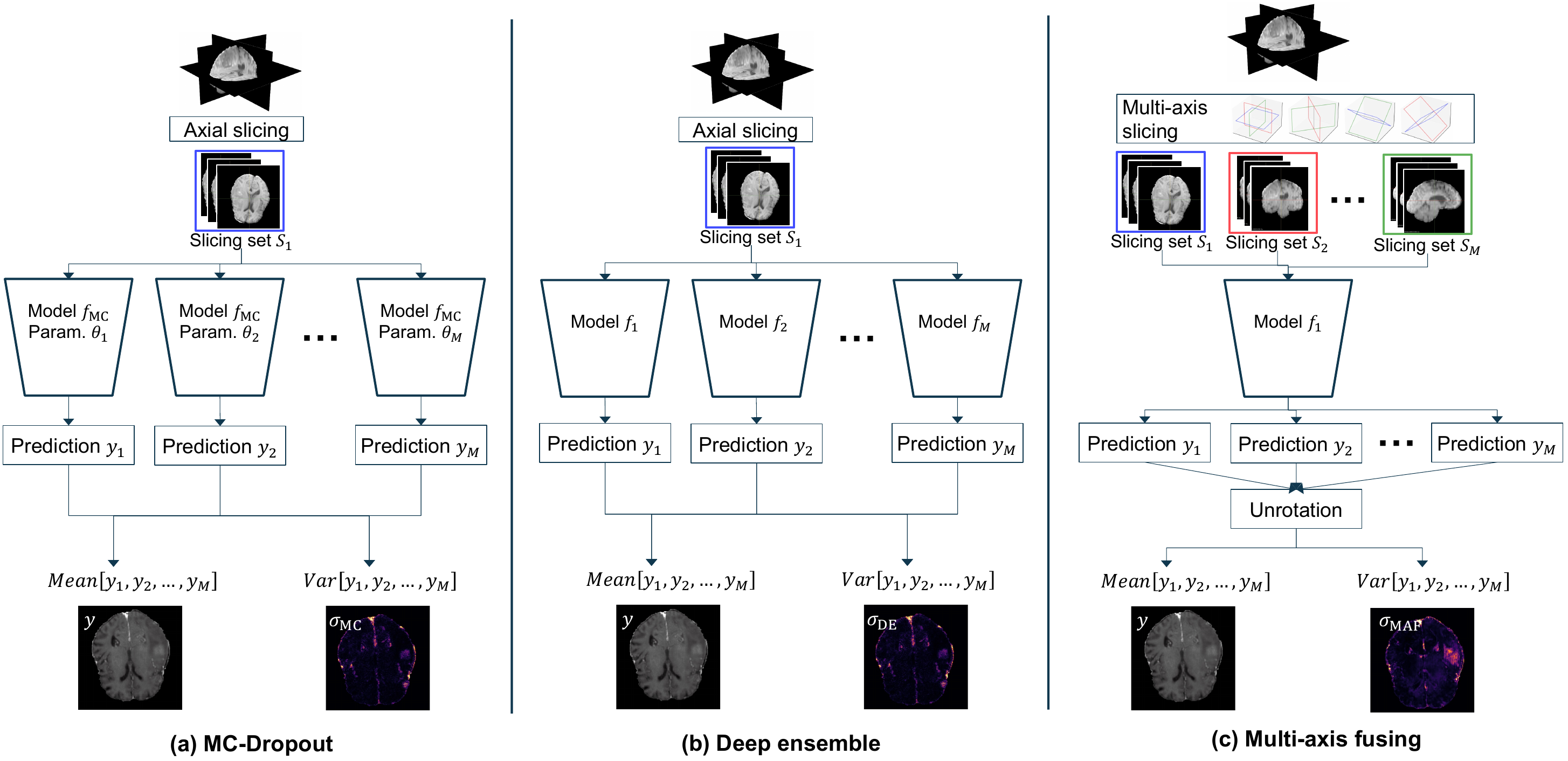}
  \caption{Comparison of three uncertainty methods. {\bf (a)} MC-Dropout employs a single model $f_{MC}$ with different parameters $\theta_1, \theta_2, \dots, \theta_M$ at inference, {\bf (b)} Deep Ensemble utilizes multiple independent models $f_1, f_2, \dots, f_M$, and {\bf (c)} Multi-axis fusion uses information from various slicing sets $S_1, S_2, \dots, S_M$ and a single model $f_1$. (Adapted from \cite{zouReviewUncertaintyEstimation2023})}
  \label{fig:1}
\end{figure*}

In this work, we investigate the efficacy of MC-Dropout, Deep Ensemble, and our method to capture uncertainty in medical image-to-image translation using a U-Net~\cite{ronnebergerUNetConvolutionalNetworks2015} style GAN model~\cite{baltruschatFRegGANKspaceLoss2023} (see Figure~\ref{fig:1}). A good uncertainty estimation aims to find a proxy for the prediction error, which is not available at the inference time. We evaluated the different uncertainty estimation techniques by comparing the mean uncertainty score with the mean absolute error (MAE) between the synthesized and ground truth images.
 
\section{Methods}
\label{sec:methods}
Let $\mathbf{V} \in \mathbb{R}^{W \times H  \times D} $ be a $3$D volume of width $W$, height $H$, and depth $D$ and let $\mathbf{I_d}$ be the $d^{th}$ $2$D slice of size $W \times H$ sliced along $\mathbf{V}$ axial plane (view). We denote the set of all slices of $\mathbf{V}$ along axial plane as $S=\{\mathbf{I}_1, \mathbf{I}_2, \dots, \mathbf{I}_D\}$. To perform image-to-image translation of $\mathbf{V}$, we synthesize a new volume $\mathbf{V}'$ of the same size as $\mathbf{V}$ by applying a trained model $f$ with parameters $\theta$ to each slice $\mathbf{I_d}$ of $\mathbf{V}$ independently. Denoting each translated slice as $\mathbf{I'_d} = f(\mathbf{I_d}, \theta)$ and translated set as $S'=\{\mathbf{I}'_1, \mathbf{I}'_2, \dots, \mathbf{I}'_D\}$, the translated volume $\mathbf{V}'$ is created by stacking the slices in $S'$ (along axial plane), i.e., $\mathbf{V}'=\text{stack}(S')$
 
\subsection{MC-Dropout}
\label{sec:dropout}
MC-Dropout \cite{galDropoutBayesianApproximation2016} uses a single model $f_{MC}$ that was trained with dropout. At inference time dropout is applied as well to stochastically sample subgraphs of the model, each uniquely described by a parameter set $\theta_1, \theta_2, \dots, \theta_M$.
For a given input $\mathbf{I_d}$, individual predictions $f_{MC}(\mathbf{I_d}, \theta_1), \dots, f_{MC}(\mathbf{I_d}, \theta_M)$ are generated for each parameter set at inference time to give $M$ samples of predictive distribution for the test case. The sample mean of these $M$ samples, i.e,
\begin{equation}\label{eq:mcdo-sample_mean}
  \mathbf{I'_d} = \frac{1}{M} \sum_{m=1}^M f_{MC}(\mathbf{I_d}, \theta_m)
\end{equation}
is used to obtain the final prediction. To estimate the associated uncertainty map $\mathbf{I_d}^{\sigma}$, the sample variance is calculated:
\begin{equation}\label{eq:mcdo-sample_var}
  \mathbf{I}^{\sigma}_d = \frac{1}{M} \sum_{m=1}^M \left( f_{MC}(\mathbf{I_d}, \theta_m) - \mathbf{I_d}' \right)^2  \text{ .}
\end{equation}
For our experiments we set $M=9$.
By stacking image slices from~\eqref{eq:mcdo-sample_mean} and uncertainty maps from~\eqref{eq:mcdo-sample_var} we obtain the final synthetic $3$D volume and its uncertainty map, respectively.

\subsection{Deep Ensemble}
\label{sec:deep-ens}
Ensembling in machine learning involves combining multiple individual models to create a more robust predictive model.
Following \cite{lakshminarayananSimpleScalablePredictive2017}, we trained $M$ models $f_1, f_2, \dots, f_M$ on the entire dataset with different random seeds to initialize the model weights, while the model architecture and the training procedure are identical.
At inference time, we computed for a given test input $\mathbf{I_d}$ the forward pass of each model $f_m$ to obtain $M$ predictive distribution samples for the test case. Similarly to~Section~\ref{sec:dropout}, we obtained $M=9$ samples using $9$ trained models. The sample mean of the individual predictions $f_1(\mathbf{I_d}),...,f_M(\mathbf{I_d})$ is computed to obtain the final synthetic image $\mathbf{I}'_d$, i.e.,
\begin{equation}\label{eq:sample_mean}
  \mathbf{I'_d} = \frac{1}{M} \sum_{m=1}^M f_{m}(\mathbf{I_d}) \text{ .}
\end{equation}
Similarly to~Section~\ref{sec:dropout}, the associated uncertainty map $\mathbf{I}_d^{\sigma}$ was obtained by computing the sample variance, and similarly stacking is used to obtain the final synthetic $3$D volume and its $3$D uncertainty map.

\subsection{Multi-Axis Fusion}
\label{sec:maf}
We adopted the multi-axis prediction fusing from~\cite{baltruschatScalingUnetSegmentation2021} to estimate the uncertainty of the synthesized image. The method is based on the observation that the segmentation of a 3D volume can be improved by fusing the predictions of multiple 2D slicing sets of the 3D volume along multiple planes (views). The slicing planes include the principal planes (axial, sagittal, and coronal) along with their corresponding oblique planes with a specific acute angle (cf. Figure~\ref{fig:1}~{\bf (c)}).
We denote each set of slices along a particular plane $m$ as $S^m=\{\mathbf{I}_1^m, \mathbf{I}_2^m, \dots, \mathbf{I}_{D_m}^m\}$.
At inference time, we evaluated each test slice in $S^m$ using a single model $f$, i.e., $\mathbf{I'^m_d} = f(\mathbf{I^m_d})$ to produce $M$ predicted slice sets, i.e., $S'^1, \dots,S'^M$.
By stacking each predicted slice set along its slicing plane, $M$ samples of predictive distribution of the translated $3$D volume are obtained, i.e., $\mathbf{V}'^m=\text{stack}(S'^m)$.
The final voxel-wise prediction is computed by averaging the predictions of the individual slicing sets, i.e.,
\begin{equation}
  \label{eq:maf}
  \mathbf{V}'_{MAF} = \frac{1}{M} \sum_{m=1}^M\mathbf{V}'^m \text{ .}
\end{equation}
The associated $3$D uncertainty map is also computed by taking the voxel-wise variance of computed $\mathbf{V}'^1,\dots,\mathbf{V}'^M$.
 
In our experiments, we used $M=9$ slicing sets. Slicing sets are obtained by slicing the 3D volume along the three principal axial, sagittal, and coronal planes. We obtained $6$ additional slicing sets with unique slicing planes and unique local context after rotating the original volume by $45$ degrees along each of the three principal axes.
 
\section{Experiments and Results}
\label{sec:exp}
We conduct a quantitative and qualitative analysis to evaluate the uncertainty estimation of the three methods. The quantitative analysis is done by correlation measurement between mean uncertainty score (MU) and mean absolute error (MAE). We qualitatively compare the uncertainty map to the mean absolute map to assess if both maps spatially correlate.
 
In our experiments, we test each method for the translation task of having T1N, T2W, and T2F sequences as input and T1C as target. We used the BraTS 2023 dataset~\cite{menzeMultimodalBrainTumor2015}. The dataset contains 1,251 exams with all four MRI scans, each with dimensions of $240 \times 240 \times 155$ voxels. We split the dataset into 1,125 for training and 126 for validation/testing. The scans have been resampled to a 1mm isotropic resolution, co-registered to a template brain, and skull-stripped. We utilize the provided segmentation masks to calculate each metric for two regions of interest (ROI), i.e., healthy and tumor. The healthy ROI is defined as the brain region excluding the tumor. The model architecture and training procedure are described in the following sections.
 
\subsection{Implementation}
Full implementation details can be found in our public code repository \url{anonymous}.
 
{\bf Model:} Due to promising results of U-net~\cite{ronnebergerUNetConvolutionalNetworks2015} style networks, we adapted an architecture from Baltruschat {\it et al.} \cite{baltruschatFRegGANKspaceLoss2023}.
Our U-net style generator consists of $8$ downsample steps where each downsampling was achieved by convolution with stride $2$ \cite{springenbergStrivingSimplicityAll2015}. Each step contains two consecutive blocks of convolution, instance normalization, and Mish\cite{misraMishSelfRegularized2020} activation function. For upsampling, we used transposed convolution. The output layer has a CeLU\cite{barronContinuouslyDifferentiableExponential2017} activation function. For the $f_{MC}$ model, we added a dropout layer after each convolution layer with a dropout rate of $0.1$. The generator is optimized with a combined loss function based on adversarial loss~\cite{maoLeastSquaresGenerative2017}, perceptual loss \cite{johnsonPerceptualLossesRealTime2016}, and frequency loss \cite{baltruschatFRegGANKspaceLoss2023}. The input to our model is a 2.5D image stack \cite{pasumarthiGenericDeepLearning2021} for each sequence, and all three sequences are combined in the channel dimension. For the 2.5D, we take two neighboring slices. Hence, in total, the input has nine channels. The output has one channel since we only predict the slice of interest for T1C.
We used a patch-wise discriminator from \cite{isolaImagetoImageTranslationConditional2018} with $5$ convolution layers (conv-layers) and added spectral normalization~\cite{miyatoSpectralNormalizationGenerative2018} to each conv-layer to stabilize the training. The discriminator is optimized with a least square loss~\cite{maoLeastSquaresGenerative2017}.
 
{\bf Dataset Preprocessing:} We used three steps for data preprocessing. First, histogram standardization was applied to all scans as proposed by Nyul {\it et al.}~\cite{nyulNewVariantsMethod2000}. The training set was used to determine intensity landmarks for each sequence, and this standardization was applied to both training and validation datasets. Secondly, MinMax intensity normalization was performed on each input and target volume set. While MinMax values were computed using only the input sequences (i.e., T1N, T2W, T2F), the normalization was applied to both input and target volumes. This was done to model the relationship of contrast uptake, resulting in an input data range of $[0, 1]$ and a possible target data range of $[0, \inf]$. Thirdly, we shift the range by linear scaling to $[-1, 1]$ and $[-1, \inf]$ for input and target, respectively.
 
{\bf Training:} To augment the training data, we sliced the scans in three basic planes which, after removing all zero slices, resulted in $502,971$ and $56,270$ slices for training and validation, respectively. Furthermore, we also applied online random cropping to $256 \times 256$ (after initially zero-padding all slices to $288 \times 288$), random horizontal flipping (probability of $p=0.5$) and random rotations in the range of $[-15, 15]$ degrees ($p=0.5$).
Our models are trained using AMSGrad optimizer with an initial learning rate of $0.0001$ and a batch size of $64$. We reduced the learning rate by a factor of two every $10$ epochs. The models are trained for $100$ epochs, where $400,000$ training images are used in each epoch.
 
\begin{table}[h]
  \caption{Pearson correlation coefficient $\rho$ and Kendall rank correlation $\tau$ between MAE and MU score in brain tumor region and healthy brain region of the image.}
  \label{tab1}
  \centering
  \begin{tabular}{l|ccc}\toprule
                               & MC-Dropout & Deep Ens. & MAF \\ \midrule \rowcolor[gray]{.95}
      Pearson $\rho_\text{healthy}$    & -0.11  & 0.38  & 0.89 \\
      Pearson $\rho_\text{tumor}$      & 0.19  & 0.45  &  0.61  \\ \midrule \rowcolor[gray]{.95}

      Kendall $\tau_\text{healthy}$    & -0.16  & 0.31  &  0.63 \\
      Kendall $\tau_\text{tumor}$      & 0.21  & 0.43  &  0.43  \\

      \bottomrule
  \end{tabular}
\end{table}
 
\subsection{Results and Discussion}
\begin{figure}[t]
  \begin{minipage}[b]{1.0\linewidth}
    \centering
    \centerline{\includegraphics[width=8.5cm]{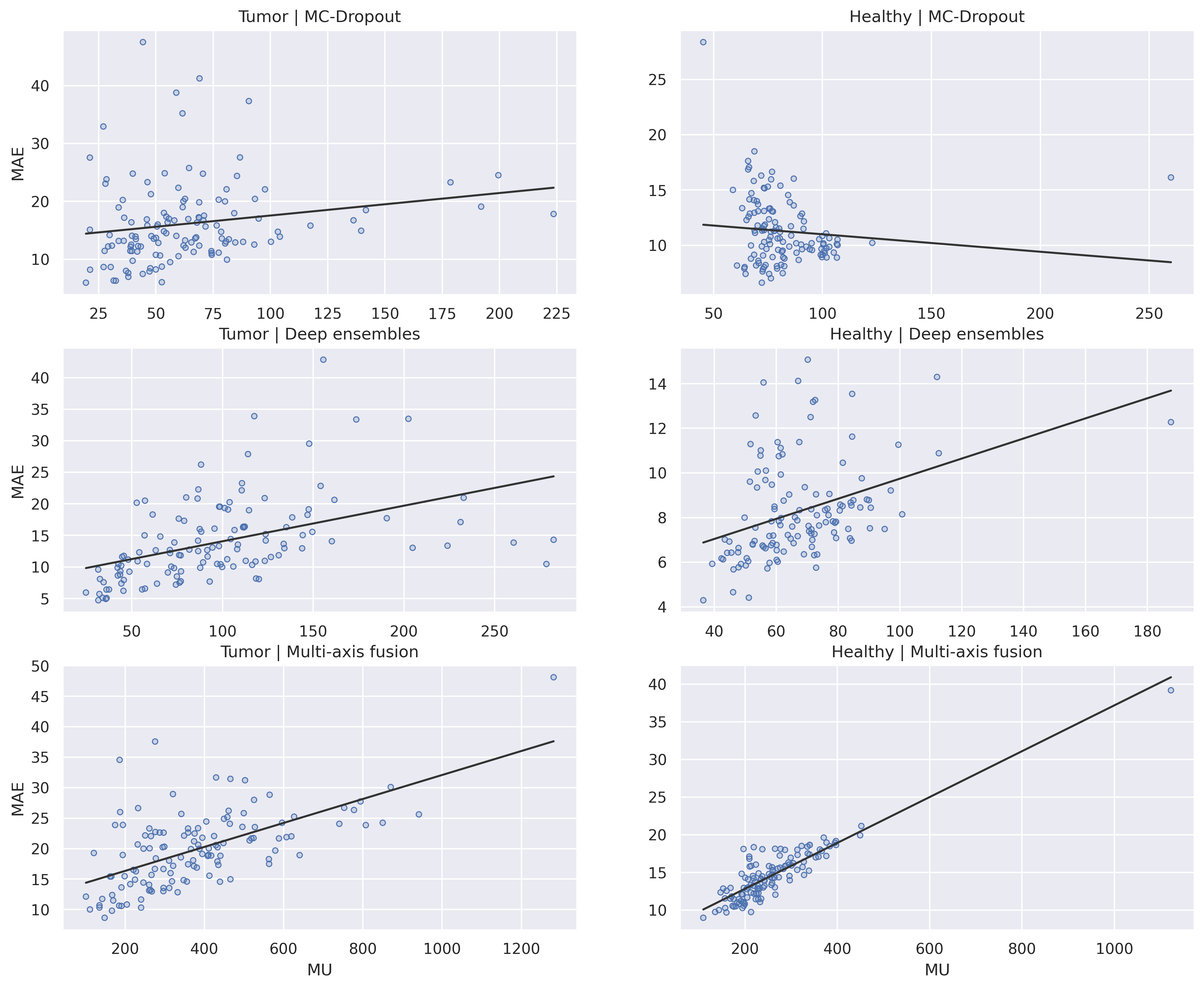}}

  \end{minipage}
  \caption{Correlation between MAE and MU score in brain tumor region (left column) and healthy brain region of the image (right column) along with the OLS linear fit (black).}
  \label{fig:2}
\end{figure}

{\noindent\bf Quantitative:} Table~\ref{tab1} shows the Pearson correlation $\rho$ and the Kendall $\tau$ rank correlation between the MAE and the MU. The values are calculated separately for the healthy ROI and the tumor ROI.
For MC-Dropout, the Pearson's correlation coefficients of both ROIs ($\rho_\text{tumor} = 0.19$ and $\rho_\text{healthy} = -0.11$) indicate that there is no linear correlation between MAE and MU. Furthermore, the associated rank correlations suggest that there is no statistically significant ordinal association between MAE and MU. This is visibly in Figure~\ref{fig:2} top row.
For Deep Ensemble, we observed a moderate Pearson correlation of MU and MAE with a higher correlation coefficient for the tumor region than for the healthy region (i.e., $\rho_\text{tumor} = 0.45$ vs. $\rho_\text{healthy} = 0.38$).
This behavior is different for MAF, where the correlation coefficient $\rho_\text{healthy} = 0.89$ is substantially greater than $\rho_\text{tumor} = 0.61$. Nevertheless, both show a high linear correlation between the mean uncertainty score and the mean absolute error, which exceed the scores of all other models. \newline

\begin{figure}[t]
\begin{minipage}[b]{1.0\linewidth}
  \centering
  \centerline{\includegraphics[width=8.5cm]{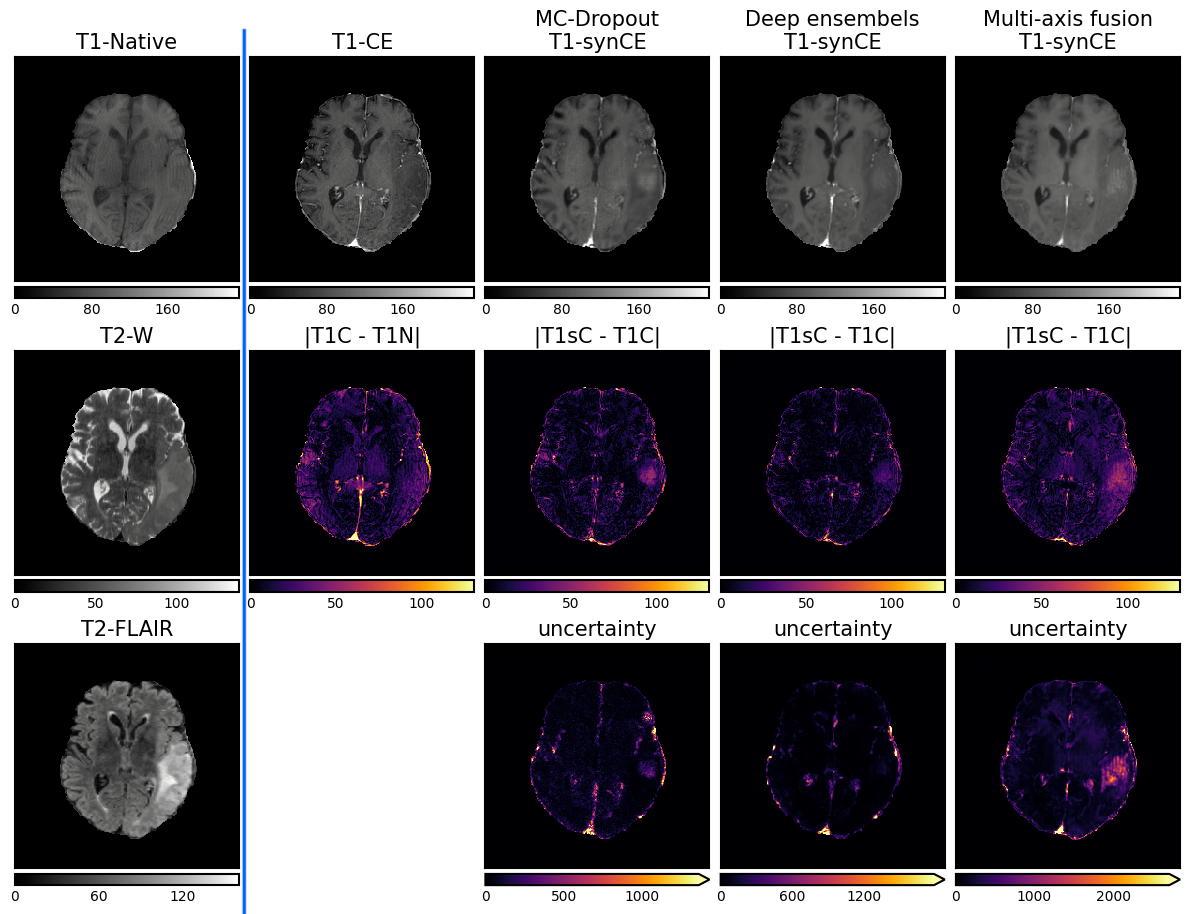}}

\end{minipage}
\caption{An exemplary absolute error and uncertainty maps. The first column shows the three input sequences: T1-Native (T1N), T2-W, T2-Flair. The second column shows: ground truth T1-CE (T1C), absolute contrast changes ($\left\lvert \text{T1C} - \text{T1N}\right\rvert $). The synthetic T1-CE image (T1sC), the error map ($\left\lvert \text{T1sC} - \text{T1C}\right\rvert $), and the uncertainty map (see Section~\ref{sec:methods}) are shown for each method in the third to the fifth column.}
\label{fig:3}
\end{figure}

{\noindent\bf Qualitative:} Figure~\ref{fig:3} shows all three methods' absolute error maps and uncertainty maps. We selected this example because all three methods synthesized a false positive enhanced tumor in the right temporal lobe. The error is less prominent for Deep Ensembles but still visible in the error map. The corresponding uncertainty map for Deep Ensemble does not capture this region; hence, it is not helpful as a proxy. MC-Dropout's uncertainty map captures this area sightly, but also some other structures not shown in the error map. The uncertainty map from our proposed method shows a strong uncertainty in the area of the false positive tumor.
 
\section{Conclusion}
\label{sec:conclusion}
We proposed a novel uncertainty estimation method, Multi-Axis Fusion, which relies on the integration of complementary information derived from multiple views on volumetric image data. The proposed approach is applied to the task of synthesizing contrast-enhanced T1-weighted images based on native T1, T2-weighted, and T2-FLAIR sequences. We compare MAF to the well-established baseline methods MC-Dropout and Deep Ensembles. Our quantitative findings indicate a strong correlation ($\rho_\text{healthy} = 0.89$, $\rho_\text{tumor} = 0.61$) between the mean absolute image synthetization error and the mean uncertainty score for our MAF method, outperforming both
MC-Dropout and Deep Ensemble models. Observing a high correlation particularly in healthy regions, suggests that the MAF confidence estimates provide a meaningful proxy for the synthetization error at the test time.

\clearpage
\bibliographystyle{IEEEbib_short}
\bibliography{references}

\end{document}